\documentclass[%
 reprint,
 aps, prl
]{revtex4-2}

\usepackage{bm,color}
\usepackage{graphicx}
\usepackage{amsmath, amssymb}
\usepackage{color}
\usepackage{braket}
\usepackage{comment}

\begin{document}
\title {Nonlinear optical response in multiband Dirac-Electron System}
\author{Keisuke Kitayama}
\affiliation{Department of Physics, University of Tokyo, Hongo, Bunkyo-ku, Tokyo 113-0033, Japan}
\author{Masao Ogata}
\affiliation{Department of Physics, University of Tokyo, Hongo, Bunkyo-ku, Tokyo 113-0033, Japan}
\affiliation{Trans-scale Quantum Science Institute, University of Tokyo, Bunkyo-ku, Tokyo 113-0033, Japan}
\begin{abstract}
We study the dc photocurrent induced by linearly polarized light in a multiband Dirac-electron system, focusing on the organic conductor $\alpha$-(BEDT-TTF)$_2$I$_3$. Utilizing perturbation theory, we predict the dependence of shift current on the frequency of light in photodriven $\alpha$-(BEDT-TTF)$_2$I$_3$. Our findings demonstrate a strong correlation between the frequency of light and both the magnitude and direction of the shift current. Furthermore, we delve into the nonperturbative effects of nonlinear optical responses using Floquet theory and demonstrate how the sign of the optical response changes with increasing light intensity. Our results unveil remarkable optical phenomena in the multiband Dirac-electron system and are anticipated to be observed in future experiments.
\end{abstract}
\maketitle


\begin{figure}[htb]
	\begin{center}
		\includegraphics[scale=0.46]{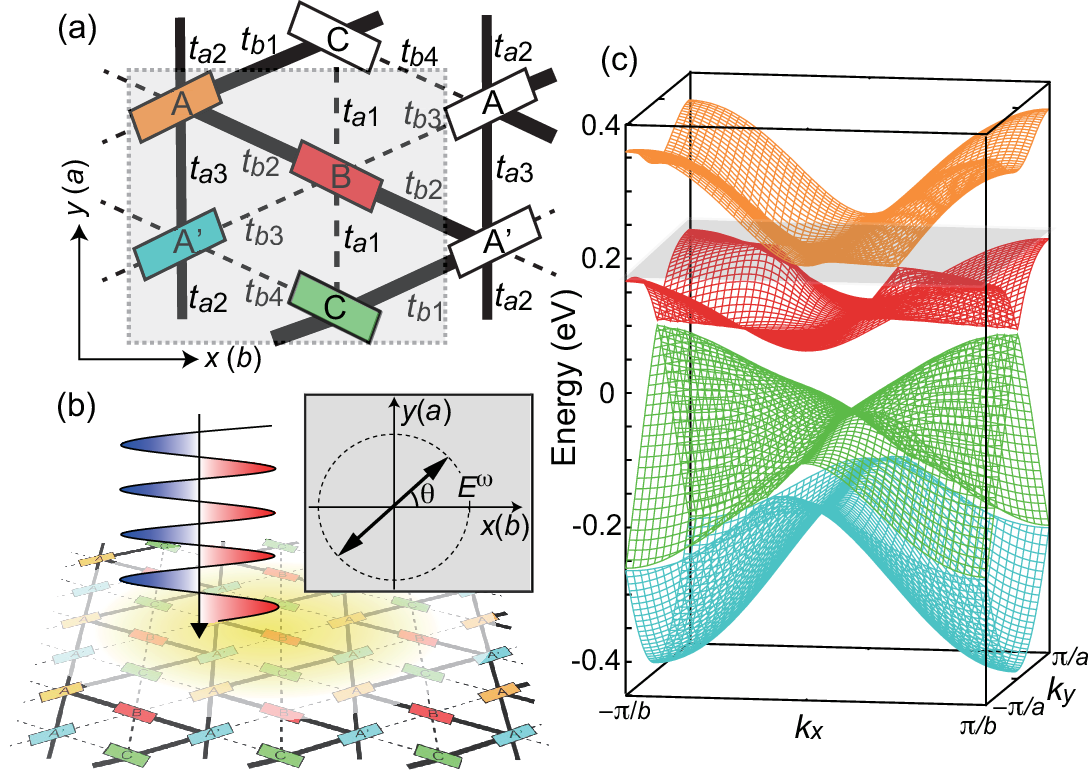}
		\caption{(a) Quasi-two-dimensional conduction layer of $\alpha$-(BEDT-TTF)$_2$I$_3$. The dashed rectangle represents a unit cell composed of four molecules (A, A$^\prime$, B, and C) and seven kinds of transfer integrals.  (b) Schematic of $\alpha$-(BEDT-TTF)$_2$I$_3$ under the irradiation with linearly polarized light. (c) Band structure of $\alpha$-(BEDT-TTF)$_2$I$_3$ obtained from a tight-binding model for two-dimensional BEDT-TTF layer.}
		\label{Fig01}
	\end{center}
\end{figure}

High-performance and cost-effective solar cell technology is of paramount importance in sustainable energy strategy. Traditional solar cells based on pn junctions encounter a fundamental limit known as the Shockley-Queisser limit, which restricts their performance~\cite{shockley1961detailed}. Therefore, it is crucial to explore alternative methods for generating photocurrent. One such promising avenue is the bulk photovoltaic effect, often referred to as the "shift current"~\cite{belinicher1980photogalvanic,von1981theory,sipe2000second}. This nonlinear optical response has the potential to generate a significantly larger dc photocurrent when compared to the pn junctions, making it an attractive alternative for efficient energy conversion~\cite{young2012first,tan2016shift,rangel2017large,ogawa2017shift,sturman2020ballistic,cook2017design}. Under these circumstances, the  shift currents have been extensively studied in gapped Dirac-electron materials such as monolayer GeS~\cite{kushnir2017ultrafast,cook2017design,kaner2020enhanced,chan2021giant}.

Historically, the shift current has been theoretically studied as a second-order response to a light electric field using perturbation theory~\cite{von1981theory, sipe2000second}. However, it is evident that higher-order terms contribute to the dc photocurrent, and nonperturbative effects have not been well understood. Recently, Morimoto and Nagaosa investigated the nonperturbative effects on the dc photocurrent in a restricted two-band system using Floquet theory, leading to a theoretical observation of saturation effect at higher light intensities~\cite{morimoto2016topological}. This saturation effect has also been experimentally demonstrated~\cite{akamatsu2021van,dong2023giant}, but a comprehensive understanding of nonlinear optical responses in multiband systems remains a subject that requires further investigation.

The organic conductor $\alpha$-(BEDT-TTF)$_2$I$_3$ offers a unique opportunity for studying nonlinear optical responses in multiband Dirac-electron systems. This compound consists of quasi-two-dimensional conduction layers composed of four BEDT-TTF molecules (A, A$^\prime$, B, C) [Fig.~\ref{Fig01}]. Under ambient pressure, this material exhibits a charge order below 135~K~\cite{takano2001charge, wojciechowski2003high}. However, when a uniaxial pressure ($P_a > 4$ kbar) is applied along the $a$ axis, the charge order dissipates~\cite{tajima2006electronic}, and the tilted Dirac cones emerge between the third and fourth bands within the set of four bands originating from the four molecules in the unit cell [Fig.~\ref{Fig01}(c)]~\cite{katayama2006pressure, kobayashi2007massless, kajita2014molecular}. Note that these four bands are situated around the Fermi level and are well-separated from the other bands, making it easier to discuss the multiband effects between these four bands~\cite{kino2006first,kitayama2021predicted,kitayama2021floquet,kitayama2022predicted}. In addition, this material is composed of large molecules and, as a result, has relatively large lattice constants ($a = 0.9187$ nm and $b = 1.0793$ nm) in comparison to graphene or transition metal dichalcogenides~\cite{mori2011charge}. This leads to the enhancement of the effects of the light electric field, which is incorporated through Peierls phases on the transfer integrals~\cite{kitayama2020predicted, kitayama2021floquet,kitayama2022predicted}. Consequently, it becomes possible to observe nonlinear effects at lower amplitude of light, so that this material holds promise as a system for studying nonperturbative effects.

In this letter, we undertake a theoretical exploration of the shift current and its nonperturbative effects in the organic compound $\alpha$-(BEDT-TTF)$_2$I$_3$ when irradiated with linearly polarized light. This serves as an example of photodriven multiband Dirac-electron systems. Through the application of perturbation theory, we predict the emergence of the shift current in photodriven $\alpha$-(BEDT-TTF)$_2$I$_3$, with the direction of the shift current being strongly contingent on the frequency of light. We uncover that this distinctive phenomenon arises from the multiband effect. Furthermore, we explore the nonperturbative effects by utilizing Floquet theory, revealing a change in the sign of the optical responses when the amplitude of light increases. We conduct a comparative analysis between the results obtained through Floquet theory and the equation proposed by Morimoto and Nagaosa, shedding light on the inherent nonlinear effects in this system.

The electronic structure of $\alpha$-(BEDT-TTF)$_2$I$_3$ can be described using a tight-binding model, given by the following equation~\cite{kitayama2021predicted,kitayama2023theoretical}:
\begin{gather}
H = \sum_{i, j}\sum_{\alpha, \beta} t_{i\alpha, j\beta} c_{i,\alpha}^{\dagger}c_{j,\beta} + \Delta\sum_{i}(c_{i,\rm{A}}^\dagger c_{i,\rm{A}} - c_{i,\rm{A}^\prime}^\dagger c_{i,\rm{A}^\prime}).
\label{Ham1}
\end{gather}
In this equation, $i, j$ refer to the unit cells, while $\alpha$ and $\beta$ label the molecular sites (A, A$^\prime$, B, and C). The symbol $c_{i,\alpha}^\dagger(c_{i,\alpha})$ represents the electron creation (annihilation) operator, and $t_{i\alpha, j\beta}$ denotes transfer integrals between neighboring sites. In Eq.~(\ref{Ham1}), we introduce a staggered site potential ($\Delta > 0$) as an order parameter for the charge order in a mean-field theory~\cite{seo2000charge}. When the charge order completely melts ($\Delta = 0$), two tilted Dirac cones emerge between the third and fourth bands. Meanwhile, when the compound is in the charge order phase ($\Delta > 0$), the tilted Dirac cones become gapped at the Dirac points, as confirmed by various experiments~\cite{wojciechowski2003high,kiswandhi2021observation}. The breaking of inversion symmetry is required to realize a nonzero dc photocurrent, so we study the latter case ($\Delta > 0$) in the present study. We set the order parameter of $\Delta = 0.01$ eV, which can be realized experimentally by applying an appropriate uniaxial pressure~\cite{kitayama2021predicted}. The transfer integrals are characterized by seven parameters as shown in Fig.~\ref{Fig01}(a), specifically: $t_{a1} = -0.038$, $t_{a2} = 0.080$, $t_{a3} = -0.018$, $t_{b1} = 0.123$, $t_{b2} = 0.146$, $t_{b3} = -0.070$, and $t_{b4} = -0.025$ eV~\cite{kobayashi2004superconductivity}. 

First, we study a dc photocurrent induced by linearly polarized light in $\alpha$-(BEDT-TTF)$_2$I$_3$ using a perturbation theory~\cite{von1981theory,sipe2000second}. We first focus on the shift-current, which is a second-order dc response to an electric field. The shift current is described by a rank-three tensor, $\sigma_{\alpha\alpha}^{\mu}$, which is given by,
\begin{eqnarray}
J^{\mu} = 2\sigma_{\alpha\alpha}^{\mu}(0,\omega,-\omega)E^{\alpha}(\omega)E^{\alpha}(-\omega).
\label{Sigma1}
\end{eqnarray}
Here, the Greek indices label spatial components, with $\mu$ belonging to the set ${x,y}$, and $J^{\mu}$ denotes $\mu$th component of a dc current density. The complex electric field $E(\omega)$ is defined as $E(t) = \mathrm{Re} [E(\omega) e^{i\omega t} + E(-\omega) e^{-i\omega t}]$, where $E(t)$ is the time-dependent electric field of light. By using the perturbation theory, the conductivity tensor in Eq.~(\ref{Sigma1}) is derived as
\begin{eqnarray}
\sigma_{\alpha\alpha}^{\mu}(0,\omega,-\omega) &=& \frac{\pi e^3}{\hbar^2 \omega^2}\sum_{m,n}\int d^2\bm{k}(f_m-f_n) |v^\mu_{mn}|^2 \nonumber \\ &\times&  \left[\frac{\partial\varphi_{mn}}{\partial k_\alpha}+A^\alpha_{m}-A^\alpha_{n}\right] \nonumber \\ &\times& \delta(\hbar\omega-\varepsilon_m+\varepsilon_n),
\label{Sigma2}
\end{eqnarray}
where $f_m$ is a Fermi-Dirac distribution function and $\varepsilon_m$ denotes the energy of the $m$th band~\cite{von1981theory,sipe2000second}. Here the quantity $\bm{A}_{n} = -i \braket{u_{n\bm{k}}|\nabla_{\bm{k}}|u_{n\bm{k}}}$ is the Berry curvature, $v^\alpha_{mn} = \braket{u_{m\bm{k}}|\partial H/\partial k_\alpha|u_{n\bm{k}}}$ denotes the matrix element of the velocity operator, $\varphi_{mn}$ represents the phase of the velocity matrix element between bands $m$ and $n$. Note that even though $\varphi_{mn}$ and $\bm{A}_{n}$ are not gauge-invariant, the conductivity in Eq.~(\ref{Sigma2}) remains gauge-invariant. For the purposes of this paper, we exclusively consider irradiation with linearly polarized light. Therefore, we do not take into account the injection current, which would be an additional contribution to the current density resulting from the breaking of time-reversal symmetry~\cite{sipe2000second,hosur2011circular,de2017quantized,chan2017photocurrents}.

\begin{figure}[tb]
	\begin{center}
		\includegraphics[scale=1.0]{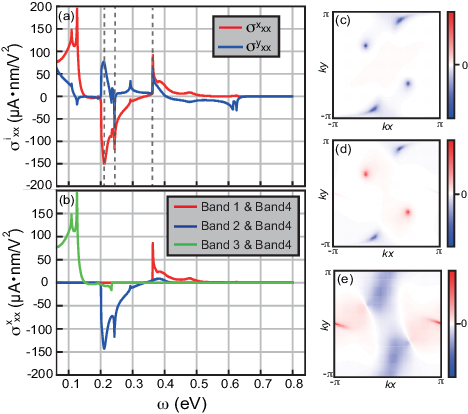}
		\caption{ (a) Shift current conductivities for $\alpha$-(BEDT-TTF)$_2$I$_3$ under the irradiation with linearly polarized light with the polarization angle of 0$^\circ$. Red and green lines indicate the $x$ and $y$ components of the shift current. (b) Contribution to the $x$ component of the shift current in photo driven $\alpha$-(BEDT-TTF)$_2$I$_3$. Blue, green, and red lines indicate the contribution from the different transitions. (c-e) The multiplication of the shift vector $S_{mn}^x$ and the velocity operator $v_{mn}^x$ in the momentum space, for various band index, (c) $m = 1$ and $n = 4$, (d) $m = 2$ and $n = 4$, and (c) $m = 3$ and $n = 4$. }
		\label{Fig02}
	\end{center}
\end{figure}

Figure~\ref{Fig02}(a) illustrates the calculated results depicting the dependence of conductivity $\sigma_{xx}^x$ and $\sigma_{xx}^y$ on the frequency of light. In this figure, we consider linearly polarized light with a polarized angle $\theta = 0^\circ$. Notably, we observe that the behavior of the conductivity $\sigma_{xx}^x$ differs significantly from that of $\sigma_{xx}^y$. As a result, the frequency of light exerts a critical impact on both the magnitude of the current density and its direction. On the other hand, it is worth mentioning that the peaks in $\sigma_{xx}^x$ and $\sigma_{xx}^y$ occurs at the same frequency as depicted in Fig.~\ref{Fig02}(a). This suggests that these peaks are attributed to the joint density of states (DOS) term, represented by $\delta(\hbar\omega - \varepsilon_m + \varepsilon_n)$.

Now, let us delve into the reason behind the significant frequency-dependent variation in the direction of the shift current. We discover that, in addition to the complex crystal structure, the multiband effect plays a crucial role in manifesting this phenomenon. To investigate this further, we compute the contribution of the shift current arising from each transition between two bands [Fig.~\ref{Fig02}(b)]. We find that the contribution from the transition between the second and fourth bands is negative, whereas the contributions from other transitions are positive. This sign disparity in contributions is attributed to the sign of the shift vector $S_{mn}^\alpha$, defined as $S_{mn}^\alpha = \partial\varphi_{mn}/\partial k_\alpha+A^\alpha_{m}-A^\alpha_{n}$. Figures~\ref{Fig02}(c-e) depict the $k$-space profiles of the multiplication of the shift vector $S_{mn}^x$ and the velocity operator $v_{mn}^x$. The shift current is obtained by multiplying the joint DOS term with $S_{mn}^xv_{mn}^x$ and integrate it across the Brillouin zone. Hence, the value of $S_{mn}^xv_{mn}^x$ at the $k$-point that satisfies $\hbar\omega = \varepsilon_m(\bm{k}) - \varepsilon_n(\bm{k})$ is the primary contribution to the shift current. As the quantity $S_{mn}^xv_{mn}^x$ is complex in momentum space, as shown in Fig.~\ref{Fig02}(c-e), the calculated contribution to the shift current from each transition, as shown in Fig.~\ref{Fig02}(b), becomes intricate when the frequency of light is altered.

Next, we explore nonperturbative effects on the dc photocurrent induced by linearly polarized light, employing the Floquet theory. The Floquet theory provides us with a method to map a time-periodically system onto a static system~\cite{aoki2014nonequilibrium, oka2019floquet, rudner2020band}. As mentioned earlier, the nonlinear optical response in a restricted two-band system has been analyzed using the Floquet theory, leading to the theoretical observation of a saturation effect at high intensity of light. Since we do not have accurate method to study the nonlinear effect at present, we employ the following approximations. First, we assume that the time-dependent Hamiltonian for $\alpha$-(BEDT-TTF)$_2$I$_3$ irradiated with linearly polarized light is given by the Peierls substitution $H_0(\bm{k} + e\bm{A}(t)/\hbar)$, where $H_0(\bm{k})$ is the Bloch Hamiltonian without the drive~\cite{oka2009photovoltaic,kitagawa2011transport,lindner2011floquet}. Second, we utilize the Keldysh formalism to study the response functions~\cite{tsuji2009nonequilibrium,aoki2014nonequilibrium}. Through the Floquet theory, the static effective Hamiltonian is given by,
\begin{eqnarray}
[H_F]_{n,m} = H_{n-m} -m\omega\delta_{n,m},
\label{Ham２}
\end{eqnarray}
where $H_n = 1/T \int dt H_0(\bm{k} + e\bm{A}(t)/\hbar) e^{in\omega t}$ is Fourier transformation of time-periodic Hamiltonian. The Floquet Hamiltonian is of infinite dimension, so we truncate it and consider a finite-dimensional matrix of 36$\times$36. The expression for the dc photocurrent in $\alpha$-(BEDT-TTF)$_2$I$_3$ is given by,
\begin{eqnarray}
J^\mu &=& \frac{ie}{2\pi\hbar}\int d^2\bm{k}\int d\nu\,\mathrm{Tr}\left[v_F G^<\right] \nonumber \\
&=& \frac{ie}{2\pi\hbar^2}\int d^2\bm{k}\int d\nu\,\mathrm{Tr}\left[ \frac{\partial H_F}{\partial k_\mu} \nonumber \right. \\
& &\times \left.\frac{1}{\nu+i\Gamma/2-H_F} \Sigma^< \frac{1}{\nu-i\Gamma/2-H_F} \right].
\label{Current1}
\end{eqnarray}
Here, $v_F$ represents the current operator under the irradiation, $G^<$ denotes the lesser Green function in the Keldysh formalism. The quantity $\Gamma$ refers to the dissipation coefficient and $\Sigma^<$ is the lesser self-energy, which is defined as,
\begin{gather}
[\Sigma^<(\nu)]_{n\nu, m\mu} = i\Gamma \left( 1 - \mathrm{tanh} \left[ \frac{\nu-\mu+m\omega}{2 k_B T} \right] \right)\delta_{n,m}\delta_{\nu,\mu}.
\label{Current2}
\end{gather}
Here, we assume that the system couples to a heat reservoir at temperature $T$ with a dissipation coefficient $\Gamma$. In the following, we set $T = 0$.

\begin{figure}[tb]
	\begin{center}
		\includegraphics[scale=1.0]{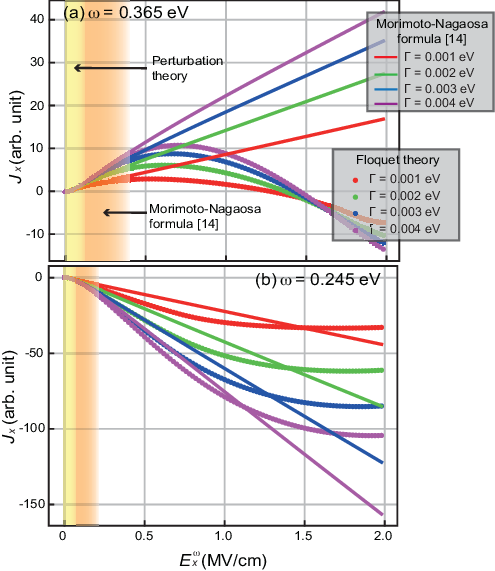}
		\caption{ Calculated nonlinear optical responses for two light frequencies (a) $\omega = 0.365$ eV and (b) $\omega = 0.245$ eV derived in Floquet theory. These frequencies correspond to the peak frequency of the shift current in Fig.~\ref{Fig02}(a). Color plots illustrate the dc photocurrent obtained using the Eq.~(\ref{Current1}), while color lines represent the dc photocurrent derived by the equation in Ref.~\cite{morimoto2016topological}. The dc photocurrents obtained using the Eq.~(\ref{Sigma2}) match those obtained from the Eq.~(\ref{Current1}) in yellow region. On the other hand, the photocurrent with the saturation factor is consistent with those obtained from Eq.~(\ref{Current1}) in orange region. }
		\label{Fig03}
	\end{center}
\end{figure}

Figure~\ref{Fig03}(a) displays the calculated dc current in photodriven $\alpha$-(BEDT-TTF)$_2$I$_3$ for various dissipation coefficients $\Gamma$. The color plots depict the results obtained using Eq.~(\ref{Current1}), while the colored lines show the outcomes calculated using Eq.~(\ref{Sigma2}) with the multiplication of a saturation factor given by 
\begin{gather}
\frac{\Gamma}{\sqrt{4e^2|v_{mn}^\mu|^2|E^\omega|^2/\omega^2 + \Gamma^2}}
\end{gather}
to the integrand [see Eq.~(15) in Ref.~\cite{morimoto2016topological}]. We verify that the photocurrent obtained using the Floquet theory is consistent with the results calculated using Eq.~(\ref{Sigma2}) when the intensity $E^\omega$ satisfies $E^\omega \lesssim 0.1$~MV/cm (yellow region). In fact, Eq.~(\ref{Sigma2}) is independent on the dissipation coefficient $\Gamma$, a feature that is only noticeable in Fig.~\ref{Fig03}(a) at the low intensity of light. On the other hand, the photocurrent calculated by Eq.~(\ref{Sigma2}) with the saturation factor match the results when $E^\omega \lesssim 0.3$~MV/cm (orange region). Notice that the saturation factor was derived in the restricted two-band system. Hence, the results in Fig.~\ref{Fig03}(a) contain the multiband effect in nonlinear optical responses. We find that the saturation factor accounts for the saturation effect at intermediate intensity of the light, but it falls short at higher intensities. Indeed, the sign change of the dc photocurrent is observed around $E^\omega \sim 1.5$ MV/cm, while the saturation factor remains positive.

Let us discuss the applicable range for the equation derived from the perturbation theory. In the perturbation theory, we develop a perturbation expansion in terms of $eE^\omega a/\hbar\omega$ where $a$ denotes the lattice constant of the system. Therefore, Eq.~(\ref{Sigma2}) is valid when the condition $eE^\omega a/\hbar\omega \ll 1$ is satisfied. As mentioned earlier, since the lattice constant in $\alpha$-(BEDT-TTF)$_2$I$_3$ is considerably larger than that in materials like graphene or transition metal dichalcogenides, the expansion parameter $eE^\omega a/\hbar\omega$ can become large, leading to a pronounced manifestation of the nonlinear effect, as depicted in Fig.~\ref{Fig03}(a). In particular, we observe a sign change of the current density as the light intensity increases. In addition, $eE^\omega a/\hbar\omega$ becomes larger when working with lower-frequency light. Indeed, the range within which Eq.~(\ref{Sigma2}) remains applicable narrows when lower-frequency light is used [Fig.~\ref{Fig03}(b)]. We observe that the applicability of Eq.~(\ref{Sigma2}) with the saturation factor also diminishes as the frequency of light decreases. In other words, in the case of a multiband system, the saturation factor discussed in Ref.~\cite{morimoto2016topological} has indeed expanded the applicable range of the equation but remains inadequate when dealing with higher $eE^\omega a/\hbar\omega$.

\begin{figure}[tb]
	\begin{center}
		\includegraphics[scale=1.0]{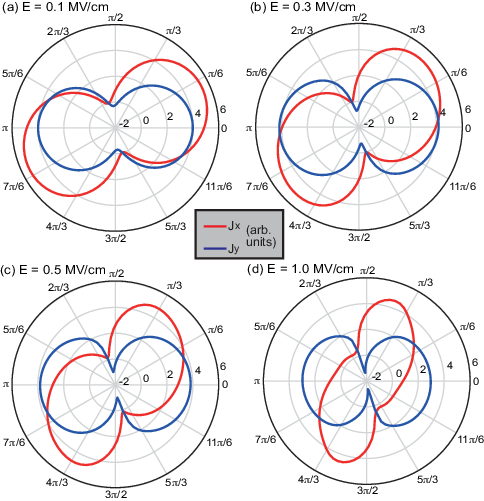}
		\caption{ The polarization angle dependence on the dc photocurrent in $\alpha$-(BEDT-TTF)$_2$I$_3$ for various intensities of light, (a) 0.1 MV/cm, (b) 0.3 MV/cm, (c) 0.5 MV/cm, and (d) 1.0 MV/cm. The frequency of light is fixed at $\omega = 0.365$ eV. Red and blue lines indicate the $x$ and $y$ components of the photocurrent. }
		\label{Fig04}
	\end{center}
\end{figure}

Finally, we investigate the dependence of the shift current on the linear polarization angle of light. In Fig.~\ref{Fig04}, the dc photocurrent density is presented as a function of the polarization angle for various amplitudes of light. Our findings reveal a significant dependence of the shift current on the polarization angle in photodriven $\alpha$-(BEDT-TTF)$_2$I$_3$. Consequently, altering the polarization angle results in changes in both the magnitude and direction of the shift current. The dependence on the polarization angle exhibits slight variations for different light amplitudes [Figs.~\ref{Fig04}(a-d)]. Moreover, it is important to emphasize that changing the polarization angle leads to a change in the sign of the $y$-component of the shift current. These results emphasize that not only the frequency but also the polarization angle are crucial parameters affecting the magnitude and direction of nonlinear optical responses.

In summary, we have predicted the nonlinear optical response in a multiband Dirac-electron system when irradiated with linearly polarized light, with a focus on $\alpha$-(BEDT-TTF)$_2$I$_3$. Employing perturbation theory, we have provided insights into the shift current's dependence on the light frequency. Our findings reveal that the direction of the shift current significantly varies with frequency, a behavior attributed to the multiband effect. Additionally, we have illustrated nonperturbative effects using Floquet theory and observed a change in the sign of the DC photocurrent as the light's amplitude increases. Our theoretical work clearly demonstrates the peculiar phenomena in multiband Dirac-electron systems and is expected to stimulate future experimental research in this area.

K.K. is grateful to S. Hasegawa, T. Joya, and H. Fukuyama for fruitful discussion. This work was supported by JSPS KAKENHI (No. 21J20856 and No. 23K03274).
\bibliography{crossresp}

\providecommand{\noopsort}[1]{}\providecommand{\singleletter}[1]{#1}%
\begin{thebibliography}{42}%
\makeatletter
\providecommand \@ifxundefined [1]{%
 \@ifx{#1\undefined}
}%
\providecommand \@ifnum [1]{%
 \ifnum #1\expandafter \@firstoftwo
 \else \expandafter \@secondoftwo
 \fi
}%
\providecommand \@ifx [1]{%
 \ifx #1\expandafter \@firstoftwo
 \else \expandafter \@secondoftwo
 \fi
}%
\providecommand \natexlab [1]{#1}%
\providecommand \enquote  [1]{``#1''}%
\providecommand \bibnamefont  [1]{#1}%
\providecommand \bibfnamefont [1]{#1}%
\providecommand \citenamefont [1]{#1}%
\providecommand \href@noop [0]{\@secondoftwo}%
\providecommand \href [0]{\begingroup \@sanitize@url \@href}%
\providecommand \@href[1]{\@@startlink{#1}\@@href}%
\providecommand \@@href[1]{\endgroup#1\@@endlink}%
\providecommand \@sanitize@url [0]{\catcode `\\12\catcode `\$12\catcode `\&12\catcode `\#12\catcode `\^12\catcode `\_12\catcode `\%12\relax}%
\providecommand \@@startlink[1]{}%
\providecommand \@@endlink[0]{}%
\providecommand \url  [0]{\begingroup\@sanitize@url \@url }%
\providecommand \@url [1]{\endgroup\@href {#1}{\urlprefix }}%
\providecommand \urlprefix  [0]{URL }%
\providecommand \Eprint [0]{\href }%
\providecommand \doibase [0]{https://doi.org/}%
\providecommand \selectlanguage [0]{\@gobble}%
\providecommand \bibinfo  [0]{\@secondoftwo}%
\providecommand \bibfield  [0]{\@secondoftwo}%
\providecommand \translation [1]{[#1]}%
\providecommand \BibitemOpen [0]{}%
\providecommand \bibitemStop [0]{}%
\providecommand \bibitemNoStop [0]{.\EOS\space}%
\providecommand \EOS [0]{\spacefactor3000\relax}%
\providecommand \BibitemShut  [1]{\csname bibitem#1\endcsname}%
\let\auto@bib@innerbib\@empty
\bibitem [{\citenamefont {Shockley}\ and\ \citenamefont {Queisser}(1961)}]{shockley1961detailed}%
  \BibitemOpen
  \bibfield  {author} {\bibinfo {author} {\bibfnamefont {W.}~\bibnamefont {Shockley}}\ and\ \bibinfo {author} {\bibfnamefont {H.~J.}\ \bibnamefont {Queisser}},\ }\bibfield  {title} {\bibinfo {title} {Detailed balance limit of efficiency of p-n junction solar cells},\ }\href@noop {} {\bibfield  {journal} {\bibinfo  {journal} {Journal of Applied Physics}\ }\textbf {\bibinfo {volume} {32}},\ \bibinfo {pages} {510} (\bibinfo {year} {1961})}\BibitemShut {NoStop}%
\bibitem [{\citenamefont {Belinicher}\ and\ \citenamefont {Sturman}(1980)}]{belinicher1980photogalvanic}%
  \BibitemOpen
  \bibfield  {author} {\bibinfo {author} {\bibfnamefont {V.~I.}\ \bibnamefont {Belinicher}}\ and\ \bibinfo {author} {\bibfnamefont {B.~I.}\ \bibnamefont {Sturman}},\ }\bibfield  {title} {\bibinfo {title} {The photogalvanic effect in media lacking a center of symmetry},\ }\href@noop {} {\bibfield  {journal} {\bibinfo  {journal} {Soviet Physics Uspekhi}\ }\textbf {\bibinfo {volume} {23}},\ \bibinfo {pages} {199} (\bibinfo {year} {1980})}\BibitemShut {NoStop}%
\bibitem [{\citenamefont {von Baltz}\ and\ \citenamefont {Kraut}(1981)}]{von1981theory}%
  \BibitemOpen
  \bibfield  {author} {\bibinfo {author} {\bibfnamefont {R.}~\bibnamefont {von Baltz}}\ and\ \bibinfo {author} {\bibfnamefont {W.}~\bibnamefont {Kraut}},\ }\bibfield  {title} {\bibinfo {title} {Theory of the bulk photovoltaic effect in pure crystals},\ }\href@noop {} {\bibfield  {journal} {\bibinfo  {journal} {Physical Review B}\ }\textbf {\bibinfo {volume} {23}},\ \bibinfo {pages} {5590} (\bibinfo {year} {1981})}\BibitemShut {NoStop}%
\bibitem [{\citenamefont {Sipe}\ and\ \citenamefont {Shkrebtii}(2000)}]{sipe2000second}%
  \BibitemOpen
  \bibfield  {author} {\bibinfo {author} {\bibfnamefont {J.~E.}\ \bibnamefont {Sipe}}\ and\ \bibinfo {author} {\bibfnamefont {A.~I.}\ \bibnamefont {Shkrebtii}},\ }\bibfield  {title} {\bibinfo {title} {Second-order optical response in semiconductors},\ }\href@noop {} {\bibfield  {journal} {\bibinfo  {journal} {Physical Review B}\ }\textbf {\bibinfo {volume} {61}},\ \bibinfo {pages} {5337} (\bibinfo {year} {2000})}\BibitemShut {NoStop}%
\bibitem [{\citenamefont {Young}\ and\ \citenamefont {Rappe}(2012)}]{young2012first}%
  \BibitemOpen
  \bibfield  {author} {\bibinfo {author} {\bibfnamefont {S.~M.}\ \bibnamefont {Young}}\ and\ \bibinfo {author} {\bibfnamefont {A.~M.}\ \bibnamefont {Rappe}},\ }\bibfield  {title} {\bibinfo {title} {First principles calculation of the shift current photovoltaic effect in ferroelectrics},\ }\href@noop {} {\bibfield  {journal} {\bibinfo  {journal} {Physical Review Letters}\ }\textbf {\bibinfo {volume} {109}},\ \bibinfo {pages} {116601} (\bibinfo {year} {2012})}\BibitemShut {NoStop}%
\bibitem [{\citenamefont {Tan}\ \emph {et~al.}(2016)\citenamefont {Tan}, \citenamefont {Zheng}, \citenamefont {Young}, \citenamefont {Wang}, \citenamefont {Liu},\ and\ \citenamefont {Rappe}}]{tan2016shift}%
  \BibitemOpen
  \bibfield  {author} {\bibinfo {author} {\bibfnamefont {L.~Z.}\ \bibnamefont {Tan}}, \bibinfo {author} {\bibfnamefont {F.}~\bibnamefont {Zheng}}, \bibinfo {author} {\bibfnamefont {S.~M.}\ \bibnamefont {Young}}, \bibinfo {author} {\bibfnamefont {F.}~\bibnamefont {Wang}}, \bibinfo {author} {\bibfnamefont {S.}~\bibnamefont {Liu}},\ and\ \bibinfo {author} {\bibfnamefont {A.~M.}\ \bibnamefont {Rappe}},\ }\bibfield  {title} {\bibinfo {title} {Shift current bulk photovoltaic effect in polar materials—hybrid and oxide perovskites and beyond},\ }\href@noop {} {\bibfield  {journal} {\bibinfo  {journal} {npj Computational Materials}\ }\textbf {\bibinfo {volume} {2}},\ \bibinfo {pages} {1} (\bibinfo {year} {2016})}\BibitemShut {NoStop}%
\bibitem [{\citenamefont {Rangel}\ \emph {et~al.}(2017)\citenamefont {Rangel}, \citenamefont {Fregoso}, \citenamefont {Mendoza}, \citenamefont {Morimoto}, \citenamefont {Moore},\ and\ \citenamefont {Neaton}}]{rangel2017large}%
  \BibitemOpen
  \bibfield  {author} {\bibinfo {author} {\bibfnamefont {T.}~\bibnamefont {Rangel}}, \bibinfo {author} {\bibfnamefont {B.~M.}\ \bibnamefont {Fregoso}}, \bibinfo {author} {\bibfnamefont {B.~S.}\ \bibnamefont {Mendoza}}, \bibinfo {author} {\bibfnamefont {T.}~\bibnamefont {Morimoto}}, \bibinfo {author} {\bibfnamefont {J.~E.}\ \bibnamefont {Moore}},\ and\ \bibinfo {author} {\bibfnamefont {J.~B.}\ \bibnamefont {Neaton}},\ }\bibfield  {title} {\bibinfo {title} {Large bulk photovoltaic effect and spontaneous polarization of single-layer monochalcogenides},\ }\href@noop {} {\bibfield  {journal} {\bibinfo  {journal} {Physical Review Letters}\ }\textbf {\bibinfo {volume} {119}},\ \bibinfo {pages} {067402} (\bibinfo {year} {2017})}\BibitemShut {NoStop}%
\bibitem [{\citenamefont {Ogawa}\ \emph {et~al.}(2017)\citenamefont {Ogawa}, \citenamefont {Sotome}, \citenamefont {Kaneko}, \citenamefont {Ogino},\ and\ \citenamefont {Tokura}}]{ogawa2017shift}%
  \BibitemOpen
  \bibfield  {author} {\bibinfo {author} {\bibfnamefont {N.}~\bibnamefont {Ogawa}}, \bibinfo {author} {\bibfnamefont {M.}~\bibnamefont {Sotome}}, \bibinfo {author} {\bibfnamefont {Y.}~\bibnamefont {Kaneko}}, \bibinfo {author} {\bibfnamefont {M.}~\bibnamefont {Ogino}},\ and\ \bibinfo {author} {\bibfnamefont {Y.}~\bibnamefont {Tokura}},\ }\bibfield  {title} {\bibinfo {title} {Shift current in the ferroelectric semiconductor sbsi},\ }\href@noop {} {\bibfield  {journal} {\bibinfo  {journal} {Physical Review B}\ }\textbf {\bibinfo {volume} {96}},\ \bibinfo {pages} {241203(R)} (\bibinfo {year} {2017})}\BibitemShut {NoStop}%
\bibitem [{\citenamefont {Sturman}(2020)}]{sturman2020ballistic}%
  \BibitemOpen
  \bibfield  {author} {\bibinfo {author} {\bibfnamefont {B.~I.}\ \bibnamefont {Sturman}},\ }\bibfield  {title} {\bibinfo {title} {Ballistic and shift currents in the bulk photovoltaic effect theory},\ }\href@noop {} {\bibfield  {journal} {\bibinfo  {journal} {Physics-Uspekhi}\ }\textbf {\bibinfo {volume} {63}},\ \bibinfo {pages} {407} (\bibinfo {year} {2020})}\BibitemShut {NoStop}%
\bibitem [{\citenamefont {Cook}\ \emph {et~al.}(2017)\citenamefont {Cook}, \citenamefont {M.~Fregoso}, \citenamefont {De~Juan}, \citenamefont {Coh},\ and\ \citenamefont {Moore}}]{cook2017design}%
  \BibitemOpen
  \bibfield  {author} {\bibinfo {author} {\bibfnamefont {A.~M.}\ \bibnamefont {Cook}}, \bibinfo {author} {\bibfnamefont {B.}~\bibnamefont {M.~Fregoso}}, \bibinfo {author} {\bibfnamefont {F.}~\bibnamefont {De~Juan}}, \bibinfo {author} {\bibfnamefont {S.}~\bibnamefont {Coh}},\ and\ \bibinfo {author} {\bibfnamefont {J.~E.}\ \bibnamefont {Moore}},\ }\bibfield  {title} {\bibinfo {title} {Design principles for shift current photovoltaics},\ }\href@noop {} {\bibfield  {journal} {\bibinfo  {journal} {Nature Communications}\ }\textbf {\bibinfo {volume} {8}},\ \bibinfo {pages} {14176} (\bibinfo {year} {2017})}\BibitemShut {NoStop}%
\bibitem [{\citenamefont {Kushnir}\ \emph {et~al.}(2017)\citenamefont {Kushnir}, \citenamefont {Wang}, \citenamefont {Fitzgerald}, \citenamefont {Koski},\ and\ \citenamefont {Titova}}]{kushnir2017ultrafast}%
  \BibitemOpen
  \bibfield  {author} {\bibinfo {author} {\bibfnamefont {K.}~\bibnamefont {Kushnir}}, \bibinfo {author} {\bibfnamefont {M.}~\bibnamefont {Wang}}, \bibinfo {author} {\bibfnamefont {P.~D.}\ \bibnamefont {Fitzgerald}}, \bibinfo {author} {\bibfnamefont {K.~J.}\ \bibnamefont {Koski}},\ and\ \bibinfo {author} {\bibfnamefont {L.~V.}\ \bibnamefont {Titova}},\ }\bibfield  {title} {\bibinfo {title} {Ultrafast zero-bias photocurrent in ges nanosheets: Promise for photovoltaics},\ }\href@noop {} {\bibfield  {journal} {\bibinfo  {journal} {ACS Energy Letters}\ }\textbf {\bibinfo {volume} {2}},\ \bibinfo {pages} {1429} (\bibinfo {year} {2017})}\BibitemShut {NoStop}%
\bibitem [{\citenamefont {Kaner}\ \emph {et~al.}(2020)\citenamefont {Kaner}, \citenamefont {Wei}, \citenamefont {Jiang}, \citenamefont {Li}, \citenamefont {Xu}, \citenamefont {Pang}, \citenamefont {Li}, \citenamefont {Yang}, \citenamefont {Jiang}, \citenamefont {Zhang} \emph {et~al.}}]{kaner2020enhanced}%
  \BibitemOpen
  \bibfield  {author} {\bibinfo {author} {\bibfnamefont {N.~T.}\ \bibnamefont {Kaner}}, \bibinfo {author} {\bibfnamefont {Y.}~\bibnamefont {Wei}}, \bibinfo {author} {\bibfnamefont {Y.}~\bibnamefont {Jiang}}, \bibinfo {author} {\bibfnamefont {W.}~\bibnamefont {Li}}, \bibinfo {author} {\bibfnamefont {X.}~\bibnamefont {Xu}}, \bibinfo {author} {\bibfnamefont {K.}~\bibnamefont {Pang}}, \bibinfo {author} {\bibfnamefont {X.}~\bibnamefont {Li}}, \bibinfo {author} {\bibfnamefont {J.}~\bibnamefont {Yang}}, \bibinfo {author} {\bibfnamefont {Y.}~\bibnamefont {Jiang}}, \bibinfo {author} {\bibfnamefont {G.}~\bibnamefont {Zhang}}, \emph {et~al.},\ }\bibfield  {title} {\bibinfo {title} {Enhanced shift currents in monolayer 2d ges and sns by strain-induced band gap engineering},\ }\href@noop {} {\bibfield  {journal} {\bibinfo  {journal} {ACS omega}\ }\textbf {\bibinfo {volume} {5}},\ \bibinfo {pages} {17207} (\bibinfo {year} {2020})}\BibitemShut {NoStop}%
\bibitem [{\citenamefont {Chan}\ \emph {et~al.}(2021)\citenamefont {Chan}, \citenamefont {Qiu}, \citenamefont {da~Jornada},\ and\ \citenamefont {Louie}}]{chan2021giant}%
  \BibitemOpen
  \bibfield  {author} {\bibinfo {author} {\bibfnamefont {Y.-H.}\ \bibnamefont {Chan}}, \bibinfo {author} {\bibfnamefont {D.~Y.}\ \bibnamefont {Qiu}}, \bibinfo {author} {\bibfnamefont {F.~H.}\ \bibnamefont {da~Jornada}},\ and\ \bibinfo {author} {\bibfnamefont {S.~G.}\ \bibnamefont {Louie}},\ }\bibfield  {title} {\bibinfo {title} {Giant exciton-enhanced shift currents and direct current conduction with subbandgap photo excitations produced by many-electron interactions},\ }\href@noop {} {\bibfield  {journal} {\bibinfo  {journal} {Proceedings of the National Academy of Sciences}\ }\textbf {\bibinfo {volume} {118}},\ \bibinfo {pages} {e1906938118} (\bibinfo {year} {2021})}\BibitemShut {NoStop}%
\bibitem [{\citenamefont {Morimoto}\ and\ \citenamefont {Nagaosa}(2016)}]{morimoto2016topological}%
  \BibitemOpen
  \bibfield  {author} {\bibinfo {author} {\bibfnamefont {T.}~\bibnamefont {Morimoto}}\ and\ \bibinfo {author} {\bibfnamefont {N.}~\bibnamefont {Nagaosa}},\ }\bibfield  {title} {\bibinfo {title} {Topological nature of nonlinear optical effects in solids},\ }\href@noop {} {\bibfield  {journal} {\bibinfo  {journal} {Science Advances}\ }\textbf {\bibinfo {volume} {2}},\ \bibinfo {pages} {e1501524} (\bibinfo {year} {2016})}\BibitemShut {NoStop}%
\bibitem [{\citenamefont {Akamatsu}\ \emph {et~al.}(2021)\citenamefont {Akamatsu}, \citenamefont {Ideue}, \citenamefont {Zhou}, \citenamefont {Dong}, \citenamefont {Kitamura}, \citenamefont {Yoshii}, \citenamefont {Yang}, \citenamefont {Onga}, \citenamefont {Nakagawa}, \citenamefont {Watanabe} \emph {et~al.}}]{akamatsu2021van}%
  \BibitemOpen
  \bibfield  {author} {\bibinfo {author} {\bibfnamefont {T.}~\bibnamefont {Akamatsu}}, \bibinfo {author} {\bibfnamefont {T.}~\bibnamefont {Ideue}}, \bibinfo {author} {\bibfnamefont {L.}~\bibnamefont {Zhou}}, \bibinfo {author} {\bibfnamefont {Y.}~\bibnamefont {Dong}}, \bibinfo {author} {\bibfnamefont {S.}~\bibnamefont {Kitamura}}, \bibinfo {author} {\bibfnamefont {M.}~\bibnamefont {Yoshii}}, \bibinfo {author} {\bibfnamefont {D.}~\bibnamefont {Yang}}, \bibinfo {author} {\bibfnamefont {M.}~\bibnamefont {Onga}}, \bibinfo {author} {\bibfnamefont {Y.}~\bibnamefont {Nakagawa}}, \bibinfo {author} {\bibfnamefont {K.}~\bibnamefont {Watanabe}}, \emph {et~al.},\ }\bibfield  {title} {\bibinfo {title} {A van der waals interface that creates in-plane polarization and a spontaneous photovoltaic effect},\ }\href@noop {} {\bibfield  {journal} {\bibinfo  {journal} {Science}\ }\textbf {\bibinfo {volume} {372}},\ \bibinfo {pages} {68} (\bibinfo {year} {2021})}\BibitemShut {NoStop}%
\bibitem [{\citenamefont {Dong}\ \emph {et~al.}(2023)\citenamefont {Dong}, \citenamefont {Yang}, \citenamefont {Yoshii}, \citenamefont {Matsuoka}, \citenamefont {Kitamura}, \citenamefont {Hasegawa}, \citenamefont {Ogawa}, \citenamefont {Morimoto}, \citenamefont {Ideue},\ and\ \citenamefont {Iwasa}}]{dong2023giant}%
  \BibitemOpen
  \bibfield  {author} {\bibinfo {author} {\bibfnamefont {Y.}~\bibnamefont {Dong}}, \bibinfo {author} {\bibfnamefont {M.-M.}\ \bibnamefont {Yang}}, \bibinfo {author} {\bibfnamefont {M.}~\bibnamefont {Yoshii}}, \bibinfo {author} {\bibfnamefont {S.}~\bibnamefont {Matsuoka}}, \bibinfo {author} {\bibfnamefont {S.}~\bibnamefont {Kitamura}}, \bibinfo {author} {\bibfnamefont {T.}~\bibnamefont {Hasegawa}}, \bibinfo {author} {\bibfnamefont {N.}~\bibnamefont {Ogawa}}, \bibinfo {author} {\bibfnamefont {T.}~\bibnamefont {Morimoto}}, \bibinfo {author} {\bibfnamefont {T.}~\bibnamefont {Ideue}},\ and\ \bibinfo {author} {\bibfnamefont {Y.}~\bibnamefont {Iwasa}},\ }\bibfield  {title} {\bibinfo {title} {Giant bulk piezophotovoltaic effect in 3r-mos2},\ }\href@noop {} {\bibfield  {journal} {\bibinfo  {journal} {Nature Nanotechnology}\ }\textbf {\bibinfo {volume} {18}},\ \bibinfo {pages} {36} (\bibinfo {year} {2023})}\BibitemShut {NoStop}%
\bibitem [{\citenamefont {Takano}\ \emph {et~al.}(2001)\citenamefont {Takano}, \citenamefont {Hiraki}, \citenamefont {Yamamoto}, \citenamefont {Nakamura},\ and\ \citenamefont {Takahashi}}]{takano2001charge}%
  \BibitemOpen
  \bibfield  {author} {\bibinfo {author} {\bibfnamefont {Y.}~\bibnamefont {Takano}}, \bibinfo {author} {\bibfnamefont {K.}~\bibnamefont {Hiraki}}, \bibinfo {author} {\bibfnamefont {H.}~\bibnamefont {Yamamoto}}, \bibinfo {author} {\bibfnamefont {T.}~\bibnamefont {Nakamura}},\ and\ \bibinfo {author} {\bibfnamefont {T.}~\bibnamefont {Takahashi}},\ }\bibfield  {title} {\bibinfo {title} {Charge disproportionation in the organic conductor, $\alpha$-({BEDT-TTF})$_2${I}$_3$},\ }\href@noop {} {\bibfield  {journal} {\bibinfo  {journal} {Journal of Physics and Chemistry of Solids}\ }\textbf {\bibinfo {volume} {62}},\ \bibinfo {pages} {393} (\bibinfo {year} {2001})}\BibitemShut {NoStop}%
\bibitem [{\citenamefont {Wojciechowski}\ \emph {et~al.}(2003)\citenamefont {Wojciechowski}, \citenamefont {Yamamoto}, \citenamefont {Yakushi}, \citenamefont {Inokuchi},\ and\ \citenamefont {Kawamoto}}]{wojciechowski2003high}%
  \BibitemOpen
  \bibfield  {author} {\bibinfo {author} {\bibfnamefont {R.}~\bibnamefont {Wojciechowski}}, \bibinfo {author} {\bibfnamefont {K.}~\bibnamefont {Yamamoto}}, \bibinfo {author} {\bibfnamefont {K.}~\bibnamefont {Yakushi}}, \bibinfo {author} {\bibfnamefont {M.}~\bibnamefont {Inokuchi}},\ and\ \bibinfo {author} {\bibfnamefont {A.}~\bibnamefont {Kawamoto}},\ }\bibfield  {title} {\bibinfo {title} {High-pressure raman study of the charge ordering in $\alpha$-({BEDT-TTF})$_2${I}$_3$},\ }\href@noop {} {\bibfield  {journal} {\bibinfo  {journal} {Physical Review B}\ }\textbf {\bibinfo {volume} {67}},\ \bibinfo {pages} {224105} (\bibinfo {year} {2003})}\BibitemShut {NoStop}%
\bibitem [{\citenamefont {Tajima}\ \emph {et~al.}(2006)\citenamefont {Tajima}, \citenamefont {Sugawara}, \citenamefont {Tamura}, \citenamefont {Nishio},\ and\ \citenamefont {Kajita}}]{tajima2006electronic}%
  \BibitemOpen
  \bibfield  {author} {\bibinfo {author} {\bibfnamefont {N.}~\bibnamefont {Tajima}}, \bibinfo {author} {\bibfnamefont {S.}~\bibnamefont {Sugawara}}, \bibinfo {author} {\bibfnamefont {M.}~\bibnamefont {Tamura}}, \bibinfo {author} {\bibfnamefont {Y.}~\bibnamefont {Nishio}},\ and\ \bibinfo {author} {\bibfnamefont {K.}~\bibnamefont {Kajita}},\ }\bibfield  {title} {\bibinfo {title} {Electronic phases in an organic conductor $\alpha$-({BEDT-TTF})$_2${I}$_3$: Ultra narrow gap semiconductor, superconductor, metal, and charge-ordered insulator},\ }\href@noop {} {\bibfield  {journal} {\bibinfo  {journal} {Journal of the Physical Society of Japan}\ }\textbf {\bibinfo {volume} {75}},\ \bibinfo {pages} {051010} (\bibinfo {year} {2006})}\BibitemShut {NoStop}%
\bibitem [{\citenamefont {Katayama}\ \emph {et~al.}(2006)\citenamefont {Katayama}, \citenamefont {Kobayashi},\ and\ \citenamefont {Suzumura}}]{katayama2006pressure}%
  \BibitemOpen
  \bibfield  {author} {\bibinfo {author} {\bibfnamefont {S.}~\bibnamefont {Katayama}}, \bibinfo {author} {\bibfnamefont {A.}~\bibnamefont {Kobayashi}},\ and\ \bibinfo {author} {\bibfnamefont {Y.}~\bibnamefont {Suzumura}},\ }\bibfield  {title} {\bibinfo {title} {Pressure-induced zero-gap semiconducting state in organic conductor $\alpha$-({BEDT-TTF})$_2${I}$_3$ salt},\ }\href@noop {} {\bibfield  {journal} {\bibinfo  {journal} {Journal of the Physical Society of Japan}\ }\textbf {\bibinfo {volume} {75}},\ \bibinfo {pages} {054705} (\bibinfo {year} {2006})}\BibitemShut {NoStop}%
\bibitem [{\citenamefont {Kobayashi}\ \emph {et~al.}(2007)\citenamefont {Kobayashi}, \citenamefont {Katayama}, \citenamefont {Suzumura},\ and\ \citenamefont {Fukuyama}}]{kobayashi2007massless}%
  \BibitemOpen
  \bibfield  {author} {\bibinfo {author} {\bibfnamefont {A.}~\bibnamefont {Kobayashi}}, \bibinfo {author} {\bibfnamefont {S.}~\bibnamefont {Katayama}}, \bibinfo {author} {\bibfnamefont {Y.}~\bibnamefont {Suzumura}},\ and\ \bibinfo {author} {\bibfnamefont {H.}~\bibnamefont {Fukuyama}},\ }\bibfield  {title} {\bibinfo {title} {Massless fermions in organic conductor},\ }\href@noop {} {\bibfield  {journal} {\bibinfo  {journal} {Journal of the Physical Society of Japan}\ }\textbf {\bibinfo {volume} {76}},\ \bibinfo {pages} {034711} (\bibinfo {year} {2007})}\BibitemShut {NoStop}%
\bibitem [{\citenamefont {Kajita}\ \emph {et~al.}(2014)\citenamefont {Kajita}, \citenamefont {Nishio}, \citenamefont {Tajima}, \citenamefont {Suzumura},\ and\ \citenamefont {Kobayashi}}]{kajita2014molecular}%
  \BibitemOpen
  \bibfield  {author} {\bibinfo {author} {\bibfnamefont {K.}~\bibnamefont {Kajita}}, \bibinfo {author} {\bibfnamefont {Y.}~\bibnamefont {Nishio}}, \bibinfo {author} {\bibfnamefont {N.}~\bibnamefont {Tajima}}, \bibinfo {author} {\bibfnamefont {Y.}~\bibnamefont {Suzumura}},\ and\ \bibinfo {author} {\bibfnamefont {A.}~\bibnamefont {Kobayashi}},\ }\bibfield  {title} {\bibinfo {title} {Molecular dirac fermion systems—theoretical and experimental approaches—},\ }\href@noop {} {\bibfield  {journal} {\bibinfo  {journal} {Journal of the Physical Society of Japan}\ }\textbf {\bibinfo {volume} {83}},\ \bibinfo {pages} {072002} (\bibinfo {year} {2014})}\BibitemShut {NoStop}%
\bibitem [{\citenamefont {Kino}\ and\ \citenamefont {Miyazaki}(2006)}]{kino2006first}%
  \BibitemOpen
  \bibfield  {author} {\bibinfo {author} {\bibfnamefont {H.}~\bibnamefont {Kino}}\ and\ \bibinfo {author} {\bibfnamefont {T.}~\bibnamefont {Miyazaki}},\ }\bibfield  {title} {\bibinfo {title} {First-principles study of electronic structure in $\alpha$-({BEDT-TTF})$_2${I}$_3$ at ambient pressure and with uniaxial strain},\ }\href@noop {} {\bibfield  {journal} {\bibinfo  {journal} {Journal of the Physical Society of Japan}\ }\textbf {\bibinfo {volume} {75}},\ \bibinfo {pages} {034704} (\bibinfo {year} {2006})}\BibitemShut {NoStop}%
\bibitem [{\citenamefont {Kitayama}\ \emph {et~al.}(2021{\natexlab{a}})\citenamefont {Kitayama}, \citenamefont {Mochizuki}, \citenamefont {Tanaka},\ and\ \citenamefont {Ogata}}]{kitayama2021predicted}%
  \BibitemOpen
  \bibfield  {author} {\bibinfo {author} {\bibfnamefont {K.}~\bibnamefont {Kitayama}}, \bibinfo {author} {\bibfnamefont {M.}~\bibnamefont {Mochizuki}}, \bibinfo {author} {\bibfnamefont {Y.}~\bibnamefont {Tanaka}},\ and\ \bibinfo {author} {\bibfnamefont {M.}~\bibnamefont {Ogata}},\ }\bibfield  {title} {\bibinfo {title} {Predicted photoinduced pair annihilation of emergent magnetic charges in the organic salt $\alpha$-({BEDT-TTF})$_2${I}$_3$ irradiated by linearly polarized light},\ }\href@noop {} {\bibfield  {journal} {\bibinfo  {journal} {Physical Review B}\ }\textbf {\bibinfo {volume} {104}},\ \bibinfo {pages} {075127} (\bibinfo {year} {2021}{\natexlab{a}})}\BibitemShut {NoStop}%
\bibitem [{\citenamefont {Kitayama}\ \emph {et~al.}(2021{\natexlab{b}})\citenamefont {Kitayama}, \citenamefont {Tanaka}, \citenamefont {Ogata},\ and\ \citenamefont {Mochizuki}}]{kitayama2021floquet}%
  \BibitemOpen
  \bibfield  {author} {\bibinfo {author} {\bibfnamefont {K.}~\bibnamefont {Kitayama}}, \bibinfo {author} {\bibfnamefont {Y.}~\bibnamefont {Tanaka}}, \bibinfo {author} {\bibfnamefont {M.}~\bibnamefont {Ogata}},\ and\ \bibinfo {author} {\bibfnamefont {M.}~\bibnamefont {Mochizuki}},\ }\bibfield  {title} {\bibinfo {title} {Floquet theory of photoinduced topological phase transitions in the organic salt $\alpha$-({BEDT-TTF})$_2${I}$_3$ irradiated with elliptically polarized light},\ }\href@noop {} {\bibfield  {journal} {\bibinfo  {journal} {Journal of the Physical Society of Japan}\ }\textbf {\bibinfo {volume} {90}},\ \bibinfo {pages} {104705} (\bibinfo {year} {2021}{\natexlab{b}})}\BibitemShut {NoStop}%
\bibitem [{\citenamefont {Kitayama}\ \emph {et~al.}(2022)\citenamefont {Kitayama}, \citenamefont {Ogata}, \citenamefont {Mochizuki},\ and\ \citenamefont {Tanaka}}]{kitayama2022predicted}%
  \BibitemOpen
  \bibfield  {author} {\bibinfo {author} {\bibfnamefont {K.}~\bibnamefont {Kitayama}}, \bibinfo {author} {\bibfnamefont {M.}~\bibnamefont {Ogata}}, \bibinfo {author} {\bibfnamefont {M.}~\bibnamefont {Mochizuki}},\ and\ \bibinfo {author} {\bibfnamefont {Y.}~\bibnamefont {Tanaka}},\ }\bibfield  {title} {\bibinfo {title} {Predicted novel type of photoinduced topological phase transition accompanied by collision and collapse of dirac-cone pair in organic salt $\alpha$-({BEDT-TTF})$_2${I}$_3$},\ }\href@noop {} {\bibfield  {journal} {\bibinfo  {journal} {Journal of the Physical Society of Japan}\ }\textbf {\bibinfo {volume} {91}},\ \bibinfo {pages} {104704} (\bibinfo {year} {2022})}\BibitemShut {NoStop}%
\bibitem [{\citenamefont {Mori}\ \emph {et~al.}(2011)\citenamefont {Mori}, \citenamefont {Usui}, \citenamefont {Sakamoto}, \citenamefont {Mizoguchi},\ and\ \citenamefont {Naito}}]{mori2011charge}%
  \BibitemOpen
  \bibfield  {author} {\bibinfo {author} {\bibfnamefont {E.}~\bibnamefont {Mori}}, \bibinfo {author} {\bibfnamefont {H.}~\bibnamefont {Usui}}, \bibinfo {author} {\bibfnamefont {H.}~\bibnamefont {Sakamoto}}, \bibinfo {author} {\bibfnamefont {K.}~\bibnamefont {Mizoguchi}},\ and\ \bibinfo {author} {\bibfnamefont {T.}~\bibnamefont {Naito}},\ }\bibfield  {title} {\bibinfo {title} {Charge distribution in the surface {BEDT-TTF} layer of $\alpha$-({BEDT-TTF})$_2${I}$_3$ at room temperature with scanning tunneling microscopy},\ }\href@noop {} {\bibfield  {journal} {\bibinfo  {journal} {Journal of the Physical Society of Japan}\ }\textbf {\bibinfo {volume} {81}},\ \bibinfo {pages} {014707} (\bibinfo {year} {2011})}\BibitemShut {NoStop}%
\bibitem [{\citenamefont {Kitayama}\ and\ \citenamefont {Mochizuki}(2020)}]{kitayama2020predicted}%
  \BibitemOpen
  \bibfield  {author} {\bibinfo {author} {\bibfnamefont {K.}~\bibnamefont {Kitayama}}\ and\ \bibinfo {author} {\bibfnamefont {M.}~\bibnamefont {Mochizuki}},\ }\bibfield  {title} {\bibinfo {title} {Predicted photoinduced topological phases in organic salt $\alpha$-({BEDT-TTF})$_2${I}$_3$},\ }\href@noop {} {\bibfield  {journal} {\bibinfo  {journal} {Physical Review Research}\ }\textbf {\bibinfo {volume} {2}},\ \bibinfo {pages} {023229} (\bibinfo {year} {2020})}\BibitemShut {NoStop}%
\bibitem [{\citenamefont {Kitayama}\ \emph {et~al.}(2023)\citenamefont {Kitayama}, \citenamefont {Ogata}, \citenamefont {Tanaka},\ and\ \citenamefont {Mochizuki}}]{kitayama2023theoretical}%
  \BibitemOpen
  \bibfield  {author} {\bibinfo {author} {\bibfnamefont {K.}~\bibnamefont {Kitayama}}, \bibinfo {author} {\bibfnamefont {M.}~\bibnamefont {Ogata}}, \bibinfo {author} {\bibfnamefont {Y.}~\bibnamefont {Tanaka}},\ and\ \bibinfo {author} {\bibfnamefont {M.}~\bibnamefont {Mochizuki}},\ }\bibfield  {title} {\bibinfo {title} {Theoretical studies of photoinduced topological phase transitions in organic salt $\alpha$-({BEDT-TTF})$_2${I}$_3$},\ }\href@noop {} {\bibfield  {journal} {\bibinfo  {journal} {JPS Conference Proceedings}\ }\textbf {\bibinfo {volume} {38}},\ \bibinfo {pages} {011150} (\bibinfo {year} {2023})}\BibitemShut {NoStop}%
\bibitem [{\citenamefont {Seo}(2000)}]{seo2000charge}%
  \BibitemOpen
  \bibfield  {author} {\bibinfo {author} {\bibfnamefont {H.}~\bibnamefont {Seo}},\ }\bibfield  {title} {\bibinfo {title} {Charge ordering in organic et compounds},\ }\href@noop {} {\bibfield  {journal} {\bibinfo  {journal} {Journal of the Physical Society of Japan}\ }\textbf {\bibinfo {volume} {69}},\ \bibinfo {pages} {805} (\bibinfo {year} {2000})}\BibitemShut {NoStop}%
\bibitem [{\citenamefont {Kiswandhi}\ and\ \citenamefont {Osada}(2021)}]{kiswandhi2021observation}%
  \BibitemOpen
  \bibfield  {author} {\bibinfo {author} {\bibfnamefont {A.}~\bibnamefont {Kiswandhi}}\ and\ \bibinfo {author} {\bibfnamefont {T.}~\bibnamefont {Osada}},\ }\bibfield  {title} {\bibinfo {title} {Observation of possible nonlinear anomalous hall effect in organic two-dimensional dirac fermion system},\ }\href@noop {} {\bibfield  {journal} {\bibinfo  {journal} {Journal of Physics: Condensed Matter}\ }\textbf {\bibinfo {volume} {34}},\ \bibinfo {pages} {105602} (\bibinfo {year} {2021})}\BibitemShut {NoStop}%
\bibitem [{\citenamefont {Kobayashi}\ \emph {et~al.}(2004)\citenamefont {Kobayashi}, \citenamefont {Katayama}, \citenamefont {Noguchi},\ and\ \citenamefont {Suzumura}}]{kobayashi2004superconductivity}%
  \BibitemOpen
  \bibfield  {author} {\bibinfo {author} {\bibfnamefont {A.}~\bibnamefont {Kobayashi}}, \bibinfo {author} {\bibfnamefont {S.}~\bibnamefont {Katayama}}, \bibinfo {author} {\bibfnamefont {K.}~\bibnamefont {Noguchi}},\ and\ \bibinfo {author} {\bibfnamefont {Y.}~\bibnamefont {Suzumura}},\ }\bibfield  {title} {\bibinfo {title} {Superconductivity in charge ordered organic conductor--$\alpha$-({ET})$_2${I}$_3$ salt--},\ }\href@noop {} {\bibfield  {journal} {\bibinfo  {journal} {Journal of the Physical Society of Japan}\ }\textbf {\bibinfo {volume} {73}},\ \bibinfo {pages} {3135} (\bibinfo {year} {2004})}\BibitemShut {NoStop}%
\bibitem [{\citenamefont {Hosur}(2011)}]{hosur2011circular}%
  \BibitemOpen
  \bibfield  {author} {\bibinfo {author} {\bibfnamefont {P.}~\bibnamefont {Hosur}},\ }\bibfield  {title} {\bibinfo {title} {Circular photogalvanic effect on topological insulator surfaces: Berry-curvature-dependent response},\ }\href@noop {} {\bibfield  {journal} {\bibinfo  {journal} {Physical Review B}\ }\textbf {\bibinfo {volume} {83}},\ \bibinfo {pages} {035309} (\bibinfo {year} {2011})}\BibitemShut {NoStop}%
\bibitem [{\citenamefont {De~Juan}\ \emph {et~al.}(2017)\citenamefont {De~Juan}, \citenamefont {Grushin}, \citenamefont {Morimoto},\ and\ \citenamefont {Moore}}]{de2017quantized}%
  \BibitemOpen
  \bibfield  {author} {\bibinfo {author} {\bibfnamefont {F.}~\bibnamefont {De~Juan}}, \bibinfo {author} {\bibfnamefont {A.~G.}\ \bibnamefont {Grushin}}, \bibinfo {author} {\bibfnamefont {T.}~\bibnamefont {Morimoto}},\ and\ \bibinfo {author} {\bibfnamefont {J.~E.}\ \bibnamefont {Moore}},\ }\bibfield  {title} {\bibinfo {title} {Quantized circular photogalvanic effect in weyl semimetals},\ }\href@noop {} {\bibfield  {journal} {\bibinfo  {journal} {Nature Communications}\ }\textbf {\bibinfo {volume} {8}},\ \bibinfo {pages} {15995} (\bibinfo {year} {2017})}\BibitemShut {NoStop}%
\bibitem [{\citenamefont {Chan}\ \emph {et~al.}(2017)\citenamefont {Chan}, \citenamefont {Lindner}, \citenamefont {Refael},\ and\ \citenamefont {Lee}}]{chan2017photocurrents}%
  \BibitemOpen
  \bibfield  {author} {\bibinfo {author} {\bibfnamefont {C.-K.}\ \bibnamefont {Chan}}, \bibinfo {author} {\bibfnamefont {N.~H.}\ \bibnamefont {Lindner}}, \bibinfo {author} {\bibfnamefont {G.}~\bibnamefont {Refael}},\ and\ \bibinfo {author} {\bibfnamefont {P.~A.}\ \bibnamefont {Lee}},\ }\bibfield  {title} {\bibinfo {title} {Photocurrents in weyl semimetals},\ }\href@noop {} {\bibfield  {journal} {\bibinfo  {journal} {Physical Review B}\ }\textbf {\bibinfo {volume} {95}},\ \bibinfo {pages} {041104(R)} (\bibinfo {year} {2017})}\BibitemShut {NoStop}%
\bibitem [{\citenamefont {Aoki}\ \emph {et~al.}(2014)\citenamefont {Aoki}, \citenamefont {Tsuji}, \citenamefont {Eckstein}, \citenamefont {Kollar}, \citenamefont {Oka},\ and\ \citenamefont {Werner}}]{aoki2014nonequilibrium}%
  \BibitemOpen
  \bibfield  {author} {\bibinfo {author} {\bibfnamefont {H.}~\bibnamefont {Aoki}}, \bibinfo {author} {\bibfnamefont {N.}~\bibnamefont {Tsuji}}, \bibinfo {author} {\bibfnamefont {M.}~\bibnamefont {Eckstein}}, \bibinfo {author} {\bibfnamefont {M.}~\bibnamefont {Kollar}}, \bibinfo {author} {\bibfnamefont {T.}~\bibnamefont {Oka}},\ and\ \bibinfo {author} {\bibfnamefont {P.}~\bibnamefont {Werner}},\ }\bibfield  {title} {\bibinfo {title} {Nonequilibrium dynamical mean-field theory and its applications},\ }\href@noop {} {\bibfield  {journal} {\bibinfo  {journal} {Reviews of Modern Physics}\ }\textbf {\bibinfo {volume} {86}},\ \bibinfo {pages} {779} (\bibinfo {year} {2014})}\BibitemShut {NoStop}%
\bibitem [{\citenamefont {Oka}\ and\ \citenamefont {Kitamura}(2019)}]{oka2019floquet}%
  \BibitemOpen
  \bibfield  {author} {\bibinfo {author} {\bibfnamefont {T.}~\bibnamefont {Oka}}\ and\ \bibinfo {author} {\bibfnamefont {S.}~\bibnamefont {Kitamura}},\ }\bibfield  {title} {\bibinfo {title} {Floquet engineering of quantum materials},\ }\href@noop {} {\bibfield  {journal} {\bibinfo  {journal} {Annual Review of Condensed Matter Physics}\ }\textbf {\bibinfo {volume} {10}},\ \bibinfo {pages} {387} (\bibinfo {year} {2019})}\BibitemShut {NoStop}%
\bibitem [{\citenamefont {Rudner}\ and\ \citenamefont {Lindner}(2020)}]{rudner2020band}%
  \BibitemOpen
  \bibfield  {author} {\bibinfo {author} {\bibfnamefont {M.~S.}\ \bibnamefont {Rudner}}\ and\ \bibinfo {author} {\bibfnamefont {N.~H.}\ \bibnamefont {Lindner}},\ }\bibfield  {title} {\bibinfo {title} {Band structure engineering and non-equilibrium dynamics in floquet topological insulators},\ }\href@noop {} {\bibfield  {journal} {\bibinfo  {journal} {Nature Reviews Physics}\ }\textbf {\bibinfo {volume} {2}},\ \bibinfo {pages} {229} (\bibinfo {year} {2020})}\BibitemShut {NoStop}%
\bibitem [{\citenamefont {Oka}\ and\ \citenamefont {Aoki}(2009)}]{oka2009photovoltaic}%
  \BibitemOpen
  \bibfield  {author} {\bibinfo {author} {\bibfnamefont {T.}~\bibnamefont {Oka}}\ and\ \bibinfo {author} {\bibfnamefont {H.}~\bibnamefont {Aoki}},\ }\bibfield  {title} {\bibinfo {title} {Photovoltaic hall effect in graphene},\ }\href@noop {} {\bibfield  {journal} {\bibinfo  {journal} {Physical Review B}\ }\textbf {\bibinfo {volume} {79}},\ \bibinfo {pages} {081406(R)} (\bibinfo {year} {2009})}\BibitemShut {NoStop}%
\bibitem [{\citenamefont {Kitagawa}\ \emph {et~al.}(2011)\citenamefont {Kitagawa}, \citenamefont {Oka}, \citenamefont {Brataas}, \citenamefont {Fu},\ and\ \citenamefont {Demler}}]{kitagawa2011transport}%
  \BibitemOpen
  \bibfield  {author} {\bibinfo {author} {\bibfnamefont {T.}~\bibnamefont {Kitagawa}}, \bibinfo {author} {\bibfnamefont {T.}~\bibnamefont {Oka}}, \bibinfo {author} {\bibfnamefont {A.}~\bibnamefont {Brataas}}, \bibinfo {author} {\bibfnamefont {L.}~\bibnamefont {Fu}},\ and\ \bibinfo {author} {\bibfnamefont {E.}~\bibnamefont {Demler}},\ }\bibfield  {title} {\bibinfo {title} {Transport properties of nonequilibrium systems under the application of light: Photoinduced quantum hall insulators without landau levels},\ }\href@noop {} {\bibfield  {journal} {\bibinfo  {journal} {Physical Review B}\ }\textbf {\bibinfo {volume} {84}},\ \bibinfo {pages} {235108} (\bibinfo {year} {2011})}\BibitemShut {NoStop}%
\bibitem [{\citenamefont {Lindner}\ \emph {et~al.}(2011)\citenamefont {Lindner}, \citenamefont {Refael},\ and\ \citenamefont {Galitski}}]{lindner2011floquet}%
  \BibitemOpen
  \bibfield  {author} {\bibinfo {author} {\bibfnamefont {N.~H.}\ \bibnamefont {Lindner}}, \bibinfo {author} {\bibfnamefont {G.}~\bibnamefont {Refael}},\ and\ \bibinfo {author} {\bibfnamefont {V.}~\bibnamefont {Galitski}},\ }\bibfield  {title} {\bibinfo {title} {Floquet topological insulator in semiconductor quantum wells},\ }\href@noop {} {\bibfield  {journal} {\bibinfo  {journal} {Nature Physics}\ }\textbf {\bibinfo {volume} {7}},\ \bibinfo {pages} {490} (\bibinfo {year} {2011})}\BibitemShut {NoStop}%
\bibitem [{\citenamefont {Tsuji}\ \emph {et~al.}(2009)\citenamefont {Tsuji}, \citenamefont {Oka},\ and\ \citenamefont {Aoki}}]{tsuji2009nonequilibrium}%
  \BibitemOpen
  \bibfield  {author} {\bibinfo {author} {\bibfnamefont {N.}~\bibnamefont {Tsuji}}, \bibinfo {author} {\bibfnamefont {T.}~\bibnamefont {Oka}},\ and\ \bibinfo {author} {\bibfnamefont {H.}~\bibnamefont {Aoki}},\ }\bibfield  {title} {\bibinfo {title} {Nonequilibrium steady state of photoexcited correlated electrons in the presence of dissipation},\ }\href@noop {} {\bibfield  {journal} {\bibinfo  {journal} {Physical Review Letters}\ }\textbf {\bibinfo {volume} {103}},\ \bibinfo {pages} {047403} (\bibinfo {year} {2009})}\BibitemShut {NoStop}%
\end{thebibliography}%
\end{document}